\documentstyle[preprint,prl,aps]{revtex}

\newcommand{\lqm}{{[\!|}}
\newcommand{\rqm}{{|\!]}}

\begin{document}
\draft

\title{EXCITON, SPINON AND SPIN WAVE MODES\\
IN AN EXACTLY SOLUBLE\\
ONE-DIMENSIONAL\\
QUANTUM MANY-BODY SYSTEM}

\author{Bill Sutherland and Rudolf A.\ R\"{o}mer}

\address{Physics Department, University of Utah, Salt Lake City, UT 84112}

\date{April 16, 1993}
\maketitle

\begin{abstract}
In this paper, we present the exact solution to a one-dimensional,
two-component, quantum many-body system in which like particles interact
with a pair potential $s(s+1)/{\rm sinh}^{2}(r)$, while unlike particles
interact with a pair potential $-s(s+1)/{\rm cosh}^{2}(r)$.
We first give a proof of integrability, then derive the coupled
equations determining the complete spectrum.  All singularities occur in the
ground state when there are equal numbers of the two components;
we give explicit results for the ground state and low-lying states in this
case. For $s>0$, the system is an antiferromagnet/insulator, with
excitations consisting of a pair-hole--pair continuum, a two-particle
continuum with gap, and excitons with gaps.
For $-1<s<0$, the system has excitations consisting of a hole-particle
continuum, and a two-spin wave continuum, both gap-less.
\end{abstract}



%
%

\section{The S-C Model}

We present the exact solution to a one-dimensional, two-component,
quantum many-body system  of considerable complexity.  The two kinds of
particles are distinguished by a quantum number $\sigma= \pm 1$, which
may be thought of as either spin or charge.
The system is defined by the Hamiltonian
\begin{equation}
H = - \sum_{1\leq j \leq N} \frac{1}{2}
       \frac{\partial^{2}}{\partial x_{j}^{2}}
    + \sum_{1\leq j < k \leq N} v_{jk}(x_{j}-x_{k}),
\end{equation}
where the pair potential is
\begin{equation}
v_{jk}(x) = s(s+1)
 \left[
  \frac{1+\sigma_{j}\sigma_{k}}{2 {\rm sinh}^{2}(x)}
  -
  \frac{1-\sigma_{j}\sigma_{k}}{2 {\rm cosh}^{2}(x)}
 \right]
\end{equation}
We assume $s\geq -1$.
We call this the S-C model, for the sinh-cosh interaction.
Thus for $s>0$, like particles repel, while unlike
particles attract.  When like particles are near, the repulsive potential
increases as $1/r^{2}$, while for large separations, both potentials decay
exponentially with a decay length we take as our length scale, and hence
unity. The potentials might usefully be thought of as a screened $1/r^2$
potential.

This system was first introduced by Calogero \cite{1}, who showed it to be
integrable. Sutherland \cite{2} soon afterward showed that the system could
be exactly solved, and gave the solution for a single component system.  He
further showed the Toda lattice to be the low-density limit, and was able
to take the classical limit to reproduce Toda's celebrated results,
identifying the particle-hole excitations of the quantum system with the
soliton-phonon modes of the classical system. Our present solution for the
two-component system exploits in a fundamental way the integrability of the
system, so we first discuss this point.

%
%

\section{Integrability}

For a classical system of $N$ one-dimensional particles, Lax \cite{3},
Moser \cite{4} and Calogero \cite{5} have shown that for certain potentials
one can find two Hermitean $N \times N$ matrices $L$ and $A$ that obey the
Lax equation ${\rm d} L / {\rm d} t = i [ A, L]$.
Thus $L$ evolves by a unitary transformation generated by $A$, and hence
 $\det [L - \omega \openone]$ is a constant of motion.
Expanding the determinant in powers of $\omega$, we find $N$ integrals of
motion, i.e.,
$
\det [ L - \omega\openone ] =
 \sum_{0 \leq j \leq N} J_{j}(-\omega)^{N-j}, j=1,\ldots,N.
$
Further, these integrals have been shown to be in involution, and thus the
system is integrable.

Under a reasonable assumption for the form of the Lax $A$ and $L$ matrices,
Calogero has shown that the most general solution to the Lax equations is
given in terms of the Jacobi elliptic function ${\rm sn}(x|m)$, and by a
suitable choice of parameters, our Hamiltonian is included.

Calogero has also demonstrated that if one replaces the classical
dynamical variables with the corresponding quantum mechanical operators,
$\det[L-\omega\openone]$ is well defined with no ordering ambiguity, and
the quantum mechanical commutator $\lqm H, \det [ L-\omega\openone] \rqm =0$.
Thus, the $J_{j}$ are still constants of motion.
Finally, Calogero showed that
$\lqm \det[L-\omega\openone], \det[L-\omega'\openone] \rqm =0$, and thus the
quantum system is also completely integrable.

To sum up the situation: Our system is completely integrable. For the
general classical system integrability tells us something concrete, namely
that the motion in terms of action-angle variables is on a torus. However,
for the general quantum system integrability seems to buy one almost nothing.
The exception is for those special cases which support scattering; i.e.,
systems which fly apart when the walls of the box are removed. In these
cases, in the distant past and future the Lax matrix $L$ approaches a
diagonal matrix with the momenta as diagonal elements, so that
$
\det[ L - \omega\openone ] = \prod_{1 \leq j \leq N} (p_{j}-\omega).
$
Thus the individual momenta $p_{j}$ are conserved in a collision, and hence,
as emphasized by Sutherland \cite{2}, the wavefunction is given
asymptotically by Bethe's ansatz. Sutherland has exploited this fact to
completely determine, in the thermodynamic limit, properties of systems
interacting by such potentials including our system. We stress that no
features of the proof of integrability are needed. All we need is to know
it to be integrable, by whatever method.

The proof of Calogero, however , is very difficult, and only briefly
sketched in the literature.  For that reason, we now offer an alternative
proof of integrability based on a method of Shastry \cite{6}.
Let us write the Lax matrices as
\begin{eqnarray}
A_{jk} &= &\delta_{jk} \sum_{l \neq j} \alpha'_{jl} +
           ( \delta_{jk} - 1) \alpha'_{jl},\\
L_{jk} &= &\delta_{jk} p_{j} + i ( 1 - \delta_{jk}) \alpha_{jk},
\end{eqnarray}
where
\begin{equation}
\alpha_{jk} =
 - s \left[ \frac{1 + \sigma_{j}\sigma_{k}}{2} {\rm coth}(x_{j}-x_{k})
          + \frac{1 - \sigma_{j}\sigma_{k}}{2} {\rm tanh}(x_{j}-x_{k})
     \right]
\end{equation}
Then if the two-body potential potential $v_{jk}$ is given by
 $v_{jk} = \alpha_{jk}^{2} + \alpha'_{jk} - s^{2}$, we find that the
quantum Lax equation $\lqm H,L \rqm = [L,A]$ is satisfied.
Here the first fancy commutator is a quantum mechanical commutator
between operators, while the second commutator is an ordered matrix
commutator, so that the equation above is really $N^{2}$ equations of the
form
$
[H,L_{jk}] = \sum_{1 \leq l \leq N}
              (L_{jl} A_{lk} - A_{jl} L_{lk})
$.
This potential, however, is exactly the potential for our system.

We now observe that the Lax $A$ matrix has the following very important
property. Defining a vector $\zeta$ with $\zeta_{j}=1$, we see
 $A \zeta = \zeta^{\dagger} A = 0$. This allows us to construct constants
of motion by $I_{n} = \zeta^{\dagger} L^{n} \zeta$, since
\begin{eqnarray}
\lqm H, I_{n} \rqm
 &= &\zeta^{\dagger} \lqm H, L_{n} \rqm \zeta \\
 &= &\zeta^{\dagger}
      \sum_{0<j<N-1}
      \left\{
       L^{j} \lqm H, L \rqm L^{N-1-j}
      \right\}
     \zeta \\
 &= &\zeta^{\dagger}
      \sum_{0<j<N-1}
      \left\{
       L^{j} [ A, L ] L^{N-1-j}
      \right\}
     \zeta \\
 &= &\zeta^{\dagger}
      \left\{
       A L^{N-1} - L^{N-1} A
      \right\}
     \zeta
  = 0
\end{eqnarray}
By Jacobi's relation for commutators, $\lqm I_{n} ,I_{m} \rqm$ is a
constant of motion, and since this is a system that supports scattering,
we see $\lqm I_{n}, I_{m} \rqm \rightarrow  0$, and hence the system
is completely integrable.

%
%

\section{The Two-Body Problem}

Having shown the system to be integrable, we then know the asymptotic
wavefunction to be of the Bethe ansatz form, and the only input needed for the
Bethe ansatz is the solution to the $2$-body problem.  Such a solution is
given in Landau and Lifshitz, {\sl Quantum Mechanics} \cite{7}.
We summarize the results below.

First, we discuss like particles. In terms of the relative coordinate
$r = x_{2} - x_{1}$, the potential is $s(s+1)/{\rm sinh}^{2}(r)$, and with
the relative momentum $k = (k_{2} - k_{1})/2$, the wave function in the
center of mass is given asymptotically as
\begin{equation}
\Psi(r) = \left\{
\begin{array}{ll}
r^{s+1}                      & r\rightarrow 0+,\\
e^{-i k r} + S(2k) e^{i k r} & r\rightarrow +\infty.
\end{array}
\right.
\end{equation}
The scattering amplitude $S(k)$ is given by
\begin{equation}
S(k) = -
\frac{\Gamma(1+ik/2) \Gamma(1+s-ik/2)}{\Gamma(1-ik/2) \Gamma(1+s+ik/2)}
\end{equation}
This scattering does not rearrange the particles.  For bosons (fermions) the
wavefunction must be (anti)symmetric, so the scattering amplitude for
transmision will be $\pm S(k)$.
In what follows, we will drop factors of $-1$ in the scattering amplitudes,
assuming that they are taken care of by either the choice of statistics of
the particles, the choice of quantum numbers as half-odd-integers, or the
choice of number of particles as even or odd.

Now, we discuss unlike particles, with the potential
$-s(s+1)/{\rm cosh}^{2}(r)$. The wave function in the center of mass is
given asymptotically as
\begin{equation}
\Psi(r) = \left\{
\begin{array}{ll}
e^{i k r} + R(2k) e^{-i k r} & r\rightarrow -\infty,\\
T(2k) e^{i k r}              & r\rightarrow +\infty.
\end{array}
\right.
\end{equation}
The reflection and transmission amplitudes are given as $R(k)=S(k)r(k)$,
$T(k)=S(k)t(k)$, where
\begin{eqnarray}
r(k) &= & \frac{\sin \pi s}{\sin \pi (s+ik/2)}, \\
t(k) &= &-\frac{\sin \pi ik/2}{\sin \pi (s+ik/2)}.
\end{eqnarray}

Also present are bound states, labeled by an index $m=1,2,\ldots,M$,
according to increasing energy. Thus the parity of the states is
$(-1)^{m-1}$, and even parity states have spin $0$, while odd parity
states have spin $1$.
Bound states appear as poles of the reflection and
transmission amplitudes, $R(k_{1}-k_{2})$ and $T(k_{1}-k_{2})$ on the
positive half of the imaginary axis, given by $k_{1,2} = k \pm i\kappa, k>0$.
The momentum and energy of such a bound state is $P = 2k$, and
$E = k^{2} - \kappa^{2}$.
{}From the particular form of the reflection and transmission amplitudes,
we find $\kappa_{m} = s+1-m$, where $m=1,2,\ldots,M$, and $M(s)$ is the
largest integer less than $s$.
There are no bound states for $0 \leq s \leq -1$. Threshhold values of $s$
are $s = 0,1,2,\dots$, and at these values, the reflection amplitude vanishes.
At the bound state poles we also find
$r(2i\kappa_{m})/t(2i\kappa_{m}) = (-1)^{m-1}$.
We call the bound states pairs.

%
%

\section{Yang-Baxter Equations}

We know the Yang-Baxter equations must hold, and we can verify this
explicitly.  For a two-component system the Yang-Baxter equations are
equivalent to $r_2 = r_3 r_1  + t_3 r_2 t_1$ and
$r_3 t_2  = r_3 t_1  + t_3 r_2 r_1$, where
$r_1 = r(k_1 -k_2), r_2 = r(k_1 -k_3 ),  r_3 = r(k_2 -k_3)$
etc.\ for $t_j$.
A degenerate situation occurs at a pole in $r_{3}$ and $t_{3}$,
when $k_{2} - k_{3} = 2i \kappa_{m}$.
There, since $r_{3}/t_{3} = (-1)^{m-1}$, the equations become
$0 = r_1  + (-1)^{m-1} r_2 t_1$ and $t_2 = t_1  + (-1)^{m-1} r_2 r_1$,
where  $r_{2,1} = r(k\pm i\kappa_{m})$, and etc. for $t_{j}$.
These relationships will be important when we calculate phase shifts.

%
%

\section{Phase Shifts}

If a particle of type $m$ passes through a particle of type $m'$, without
reflection, then we have a scattering amplitude
$\exp[-i\theta_{mm'}(k_1 - k_2)]$ , and a corresponding phase shift
$\theta_{mm'}(k)$.
These phase shifts are at the heart of the Bethe ansatz. Let us label the
unbound particle by $m = 0$. Then we have found
\begin{equation}
\theta_{00}(k) =
i \log \left[
\frac{ \Gamma(1+ik/2)\Gamma(1+s-ik/2) }{ \Gamma(1-ik/2)\Gamma(1+s+ik/2) }
\right]
\end{equation}
As we explained, we will not include factors of $-1$ in the scattering
amplitudes, so $\theta_{00}(0) = 0$.
In general $\theta_{mm'}(k) = -\theta_{mm'}(-k) = \theta_{m'm}(k)$,
and we will find that we always have $\theta_{mm'}(0) = 0$.

Now, consider the scattering of a particle $k_1$ on a pair of two
particles with momenta $k_2 \pm i \kappa_m$.  Let $k = k_1 - k_2$.
Then using the degenerate Yang-Baxter equations, we find for the
scattering amplitude
\begin{equation}
S(k + i\kappa_m) S(k - i\kappa_m) t(k + i\kappa_m) = \exp[-i\theta_{0m}(k)].
\end{equation}
Using the explicit forms, we can verify that $\theta_{0m}(k)$ is real
for $k$ real.

Finally, we view the scattering of a pair from a pair as the scattering of two
particles with momenta $k_1 \pm i \kappa_m$ from a pair with
$k_2 \pm i \kappa_{m'}$.  This gives us a net phase shift
$
\theta_{mm'}(k) = \theta_{0m'}(k - i\kappa_m) + \theta_{0m'}(k + i\kappa_m).
$
Again, using the explicit forms, we can verify that $\theta_{mm'}(k)$
is real for $k$ real, and symmetric in $m, m'$.

To summarize: We have $N_\uparrow$ particles with $\sigma= +1$, and
$N_\downarrow$ with $\sigma= -1$, for a total of
$N = N_\uparrow + N_\downarrow$.
Let us assume $N_\uparrow \geq N_\downarrow$. Further, pairs of up-down spins
bind into a variety of bound states, or pairs, labeled by $m$,
$m = 1,\ldots,M(s)$. Let there be $N_m$ of each type. Then the number of
unbound particles is
$
N_0 = N - 2 \sum_{1 \leq m \leq M} N_m,
$
We will call these simply particles from now on.
They would correspond to spinons/ions in the spin/charge picture.
Of these particles, we have $N_{-1}$ with spin down; let us call them
spin waves. Clearly
$
N_{-1} = N_\downarrow - \sum_{1 \leq m \leq M} N_m,
$
and $N_{-1} \leq N_0/2$.

We still must treat the dynamics of the spin waves, but since they are not
``real'' particles, but only correlations in the quantum numbers of particles,
they have no momentum or energy directly.  Thus, defining
\begin{equation}
\eta_m = \left\{
\begin{array}{ll}
0,              & m=-1,\\
1,              & m=0,\\
2,              & m=1,2,\ldots,M(s)
\end{array}
\right.
\end{equation}
then we can write the momentum and energy as
\begin{eqnarray}
P &= &\sum_{-1 \leq m \leq M} \eta_m \sum_{k_m}k_m,\\
E &= &\frac{1}{2}
      \sum_{-1 \leq m \leq M} \eta_m \sum_{k_m}k_{m}^2
     -\sum_{1 \leq m \leq M} N_m \kappa_{m}^2,\\
\end{eqnarray}

%
%

\section{Spin Waves}

Since particle-pair and pair-pair pass through one another with only a
phase shift and no reflection, their interaction is in some sense trivial.
However, particles do scatter from particles with reflection, and their
interaction is not trivial.
We write the asymptotic wavefunction explicitly in the Bethe ansatz
form, and for now consider only the $N_0$ particles. We use the spin language,
so $\sigma_z(j) = \pm 1$ according to whether the $j$th particle in the
ordering  $x_1<...<x_{N_0}$ has spin up or down.
A choice for all $\sigma_z(j)$ we denote simply by $\sigma$. Then
asymptotically the wave function is given by
\begin{equation}
\Psi(x|\sigma) \rightarrow
\sum_{P} A(P|\sigma) \exp\left[ i \sum_{1 \leq j \leq N_0} x_j k_{Pj}
                         \right].
\end{equation}
The summation is over all the $N_0 !$ permutations of the momenta.
We arrange the $A(P|\sigma)$ for fixed $P$ as a column vector $\xi(P)$.
Then the Yang-Baxter equations ensure that we can find a consistent set of
amplitudes $A(P|\sigma)$, by finding the simultaneous eigenvector of the
$N_0$ equations
\begin{equation}
e^{i k_j L}
\prod_{1 \leq n \leq N_0} S(k_j-k_n)
 X_{j,j-1} \cdots X_{j,1}
 X_{j,N_0} \cdots X_{j,j+1}
 \xi(I) = \xi(I)
\end{equation}
In this equation, the $X_{j,n}$ are operators given as
\begin{eqnarray}
X_{j,n} &= &
 \frac{1+t(k)}{2} \openone
+\frac{1-t(k)}{2} \sigma_z(j)\sigma_z(n) \nonumber \\
        &\mbox{ } &
+\frac{r(k)}{2} \left[ \sigma_x(j)\sigma_x(n) + \sigma_y(j)\sigma_y(n)
                \right],
\end{eqnarray}
where  $k = k_{P_j} - k_{P_n}$.

These eigenvalue equations can in turn be solved by a Bethe ansatz for the
$N-1$ overturned spins --- the spin waves --- on a lattice of $N_0$ particles.
These equations can be solved either: (i) directly, by the methods of
Yang \cite{8}; (ii) with commuting transfer matrices, by the methods of
Baxter \cite{9}; or (iii) by quantum inverse scattering methods of Faddeev
and Takhtajan \cite{10}. We are not aware that these equations have appeared
before in the solution of a quantum many-body problem, although the
low-density case has often appeared, for instance first in Yang's original
solution for $\delta$-function fermions.

The solution is sufficiently technical that we postpone discussion to a later
publication. However, the result has many interesting physical consequences,
and that is what we want to discuss in this letter. One finds for the
eigenvalues of the previous equations,
\begin{equation}
e^{i k_j L}
\prod_{1 \leq n \leq N_0} S(k_j-k_n)
\prod_{1 \leq q \leq N_{-1}}
\frac{\sin\pi[s-i(k_j-\lambda_q)]/2}{\sin\pi[s+i(k_j-\lambda_q)]/2}
= 1
\end{equation}
In this equation, the $\lambda$'s are the momenta of the spin waves, and are
determined from the equation
\begin{equation}
\prod_{1 \leq q \leq N_{-1}}
 \frac{\sin\pi[s+i(\lambda_j-\lambda_q)/2]}
      {\sin\pi[s-i(\lambda_j-\lambda_q)/2]}
\prod_{1 \leq n \leq N_0}
 \frac{\sin\pi[s-i(\lambda_j-k_n)]/2}
      {\sin\pi[s+i(\lambda_j-k_n)]/2}
= 1
\end{equation}
We now have our two final phase shifts, for particle-spin wave and spin
wave-spin wave
scattering:
\begin{eqnarray}
\theta_{0,-1}(k)
 &= &i \log\left[
 \frac{\sin\pi[s-ik]/2}{\sin\pi[s-ik]/2}
       \right],\\
\theta_{-1,-1}(k)
 &= &i \log\left[
 \frac{\sin\pi[s+ik/2]}{\sin\pi[s-ik/2]}
       \right].
\end{eqnarray}
As noted, there is no phase shift for spin wave-pair scattering.

At the threshhold values $s = {\rm integer}$, the spin wave modes uncouple
completely from the particles, and thus from the system, since they
contribute no energy or momentum directly.
In this case, we have the very high degeneracies found in the $1/r^2$
lattice systems \cite{11,12}, and for the same reason --- an absence of
reflection.

%
%

\section{The Solution}

Let us now impose periodic boundary conditions and take any particle,
pair or spin wave around a ring of large circumference. Along the
way, it suffers a phase change as it scatters from every other particle,
pair or spin wave, plus a phase change of $PL$, where $P = \eta k$ is its
own momentum.
Periodicity requires that this phase change be an integer multiple of $2 \pi$,
the integer being the quantum number.  We write this statement as coupled
equations in a rather symbolic form:
\begin{equation}
L \eta_m k_m = 2\pi I_m(k_m)
 + \sum_{-1 \leq m' \leq M} \sum_{k'_{m'}}
    \theta_{m,m'}(k_m - k_{m'}), \quad m= -1, 0, 1, \ldots, M.
\end{equation}
Here the $I_m(k_m)$ are the quantum numbers, the only subtlety being that
for the spin waves, $I_{-1}$ ranges only over $1,\ldots,N_0$.

%
%

\section{Results for Zero Temperature and Zero Spin/Charge}

In this letter, we give explicit results for the ground state and low-lying
states when $N_\uparrow = N_\downarrow$, which we call the spin/charge
zero-sector.  This certainly is the most interesting case, since all
singularities in the $(N_\uparrow, N_\downarrow )$ ground state
phase diagram occur for $N_\uparrow = N_\downarrow$. In fact, as we shall
see, for $s>0$, the chemical potential has a discontinuity across the line
$N_\uparrow = N_\downarrow$, and thus the system is an
antiferromagnet/insulator, although not of the Neel/Mott type.
For $-1<s<0$, there is a weak singularity at $N_\uparrow = N_\downarrow$,
without a discontinuity in the chemical potential.

For $s>0$, which we call the attractive case, the ground state consists of
a spin fluid of type $m=1$, and thus spin $0$.
This is the bound state with lowest binding energy, when
$\kappa=s$, and so $P(k) = 2k$ and $E(k) = k^2 - s^2$.
In the ground state, the $k$'s for the pairs distribute themselves densely
with a density $\rho(k)$, between limits $\pm B$, normalized so that
\begin{equation}
N_{1}/L = \int_{-B}^{B} \rho(k) {\rm d}k = N/2L.
\end{equation}
The energy and momentum are given by
\begin{eqnarray}
P/L &= & 2 \int_{-B}^{B} \rho(k) k {\rm d}k = 0,\\
E/L &= &   \int_{-B}^{B} \rho(k) k^2 {\rm d}k - s^2 N_1/L.
\end{eqnarray}
The integral equation which determines $\rho(k)$ is
\begin{equation}
1/\pi = \rho(k) + \frac{1}{2\pi} \int_{-B}^{B} \theta'_{11}(k-k')\rho(k')
 {\rm d}k'.
\end{equation}
The kernel of the equation, $\theta_{11}'(k)$, is the derivative of the
phase shift.  In figure (\ref{fig-1}) we show $E_0/L$ versus $N/L$ for
selected values of $s = 1/2, 1, 3/2$.

	Having determined the ground state properties of the system, we now
determine the low-energy excited states.  They are given by the following:
(i) Remove a pair from the ground state distribution, and place it outside the
limits; we call this creating a pair-hole and a pair, and it gives a two
parameter continuum.
(ii)  Break a pair, to give two particles, one spin up and the other spin
down;  this also gives a two parameter continuum.
(iii)  Excite a pair into a higher energy bound state, if allowed; these we
call excitons, and they have single parameter dispersion relations.
(Away from the zero-sector, we can have in addition spin waves. These will be
important for $s<0$.)

By the techniques of Yang and Yang \cite{13}, the dispersion relations are
given parametrically by
\begin{eqnarray}
\Delta P &=
 & \sum_{m} \left[
    \eta_m k_m - \int_{-B}^{B} \theta_{m1}(k_m-k)\rho(k){\rm d}k
   \right],\\
\Delta E &=
 & \sum_{m} \left[
    \frac{\eta_m k_m^2}{2} -
    \frac{1}{2\pi} \int_{-B}^{B} \theta'_{m1}(k_m-k)\epsilon(k){\rm d}k
   \right].
\end{eqnarray}
Here $\epsilon(k)$ is the solution to the integral equation
\begin{equation}
k^2 - s^2 = \mu_1 =
\epsilon(k) + \frac{1}{2\pi}
 \int_{-B}^{B} \theta'_{11}(k-k')\epsilon(k'){\rm d}k'.
\end{equation}
The chemical potential $\mu_1$ is the chemical potential for pairs, given by
$\partial E_0 / \partial N_1$.
The results are shown in figure (\ref{fig-2}) for
$s = 3/2$, $B = 3/2$, $d = N/L = 0.943$, $E_0/L= -0.691$, $\mu_1 = 1.215$.
The gap for the creation of two particles is $\Delta E = 1.170$,
and is equal to the discontinuity of the chemical potential across the line
$N_\uparrow = N_\downarrow$.
The exciton with $m=2$ is the only exciton allowed at this value of $s$,
and has a gap of $\Delta E = 1.017$.

For $0>s>-1$, in the zero-sector, we have two coupled equations for $N$
particles and $N/2$ spin waves. However, in the zero-sector, the limits of the
spin wave distribution are $\pm\infty$. Thus we can solve by Fourier transforms
for the spin wave distribution in terms of the particle distribution, and then
substitute this into the particle equation, giving us a single integral
equation for the distribution of particles $\rho(k)$:
\begin{equation}
\frac{1}{2\pi} = \rho(k) + \frac{1}{2\pi}
 \int_{-B}^{B} \theta'(k-k')\rho(k'){\rm d}k'.
\end{equation}
Here the kernel $\theta'(k)$ is given as
\begin{equation}
\theta'(k)= \theta'_{00}(k) -
 2 \int_{-\infty}^{\infty} {\rm d}t e^{i k t}
\frac{{\rm sinh}^3 t(1+s) {\rm cosh} ts}{{\rm sinh}^3 t}.
\end{equation}
In figure (\ref{fig-1}), we show $E_0/L$ versus $N/L$ for $s = -1/2$.
The excited states in the zero-sector are given by:
(i) Remove a particle from the ground state distribution, and place it
outside the limits; we call this creating a hole and a particle, and it
gives a two parameter continuum.
(ii) Remove a spin wave from the ground state distribution, and place it on
the line with imaginary part equal to $i$; we call this creating two spin
waves,
one with spin up and the other with spin down. It gives a two parameter
continuum, familiar from the Heisenberg-Ising model. The results are shown
in figure (\ref{fig-3}), for $s=-1/2$, $B = 1$, $d = N/L = 0.600$,
$E_0/L = 0.094$, $\mu = 0.374$.

Finally, we remark that all thermodynamics can be explicitly calculated,
since there are no ambiguities with counting states, or difficulties with
strings of length greater than two.

\acknowledgements
We would like to thank Sriram Shastry for his interest, insight and
encouragement.  In addition, we would like to thank the creators of
{\sl Mathematica$^{{\sc\tiny TM}}$} for the {\tt LogGamma[]} and
{\tt PolyGamma[]} functions, without which this investigation would be
where it was fifteen years ago.


\begin{figure}
\caption{Ground state energy per unit length $E_0/L$ versus density $N/L$ for
$s = -1/2, 1/2, 1, 3/2$.}
\label{fig-1}
\end{figure}

\begin{figure}
\caption{Energy above the ground state energy versus momentum (dispersion
relations) for the low lying excitations when $s = 3/2$ and density
$N/L = 0.943$.}
\label{fig-2}
\end{figure}

\begin{figure}
\caption{Energy above the ground state energy versus momentum (dispersion
relations) for the low lying excitations when $s = -1/2$ and density
$N/L =0.600$.}
\label{fig-3}
\end{figure}

\end{document}